\documentclass[doublecol]{epl2}
\usepackage{graphicx}
\usepackage{subfigure}
\usepackage{color}
\usepackage{amsmath}
\usepackage{amssymb}

\usepackage{dcolumn}
\usepackage{bm}

\title{Effects of Spatial Dispersion on the Casimir Force between Graphene Sheets}

\author{D. Drosdoff\inst{1}, A. D. Phan\inst{1}, L. M. Woods\inst{1}, I. V. Bondarev\inst{2}, and J. F. Dobson\inst{3}}
\institute{                    
  \inst{1} Department of Physics, University of South Florida, Tampa FL 33620

  \inst{2} Department of Physics, North Carolina Central University, Durham NC 27707
  
  \inst{3} Micro and Nanotechnology Centre, Griffith University, Nathan, Queensland 4111, Australia
}
\pacs{42.50.Lc}{Quantum fluctuations, quantum noise, and quantum jumps}
\pacs{72.80.Vp}{Electronic transport in graphene }
\pacs{73.22.Pr}{	Electronic structure of graphene}

\abstract{The Casimir force between graphene sheets is investigated with emphasis on the effect from spatial dispersion using a combination of factors, such as a nonzero chemical potential and an induced energy gap.  We distinguish between two regimes for the interaction - $T=0$ $K$ and $T\neq 0$ $K$. It is found that the quantum mechanical interaction ($T=0$ $K$) retains its distance dependence regardless of the inclusion of dispersion. The spatial dispersion from the finite temperature Casimir force is found to contribute for the most part from $n=0$ Matsubara term. These effects become important as graphene is tailored to become a poor conductor by inducing a band gap.}

\begin{document}

\maketitle

\section{Introduction \label{INTRO}}

Long-ranged dispersive interactions, such as the Casimir\cite{Casimir:1948} force or the van der Waals\cite{London:1930} force, originate from electromagnetic field fluctuations. They are present between all types of objects at any distance, regardless of whether there are permanent electric and/or magnetic moments. Despite their universal nature, the dimensionality, dielectric and magnetic response properties, and geometry of the interacting systems can influence the dispersive force in profound ways, thus offering possibilities to tailor the sign, magnitude, and distance dependences of the interaction\cite{Abrikosov:1975,Rodriguez:2011}. 

Single layers of graphite have been recently isolated \cite{Novoselov:2004,Novoselov:2005}, and their dispersive forces present not only challenging theoretical problems, but they are also important for graphene based devices\cite{Lin:2010}, \cite{Stoller:2008}. The electromagnetic fluctuation forces in graphene have been considered both at low and high temperature regimes as well as with the inclusion of a chemical potential or a band gap \cite{Dobson:2011,Drosdoff:2011,Sarabadani:2011, Svetovoy:2011, Fialkovsky:2011, Sernelius:2011, Drosdoff:2010, GomezSantos:2009,Dobson:2006}. A distance power law has been obtained for the graphene/graphene Casimir force at the quantum limit, where the force per unit area($F$) goes as the inverse distance($d$) to the fourth power, $F\sim d^{-4}$. At the same time, due to the graphene gapless electronic structure, the fluctuation forces are thermal in nature at room temperature. This is in stark contrast to the Casimir forces in most matter whereupon quantum fluctuation effects are dominant even at relatively high temperatures.  

Concerning Casimir interactions in general, the inclusion of spatial dispersion  in the response properties of the materials has received significant attention recently, since it was shown that the dispersion may play an important role  \cite{Sernelius:2006,Esquivel:2006}. In this work we investigate to what extent spatial dispersion, in combination with modifications in the graphene electronic structure through a nonzero chemical potential or an induced energy gap, affect the graphene-graphene Casimir interaction.
In practice, a finite chemical potential is induced by doping\cite{Brenner:2012,Lin:2012} or by the application of external fields\cite{Novoselov:2004}.  Energy band gaps on the other hand can be brought about by the growth of graphene on certain substrates\cite{Zhou:2007}, via adsorption\cite{Yavari:2010} or hydrogenation\cite{Haberer:2010}. The response properties of graphene for the calculations here are taken into account via the conductivity obtained from the linear response to the electromagnetic field\cite{Falkovsky:2007}, from which both the small and large spatial dispersion forces are investigated.

\section{Graphene Conductivity\label{GC}}

An important factor in the Casimir force is the dielectric response properties of the materials involved. For the graphene/graphene system, these properties are described via the optical conductivity calculated within the Kubo\cite{Kubo:1957} formalism. Consider the case when there is a band gap $\Delta$ induced at the graphene Fermi level. Near the K-point of the Brillouin zone, assuming small wave vector excitations, the Hamiltonian can be written as 
 \begin{equation}
H=
\left(
\begin{array}{cc}
\Delta & v_0(p_x-ip_y) \\
v_0(p_x+ip_y) & -\Delta \\
\end{array}
\right),
\label{GC2}
\end{equation} 
where ${\bf p}$ is the two-dimensional momentum operator and $v_0=3\gamma_0 a/2\approx 10^6 m/s$ ($a=0.142\ nm$ is the CC distance  and $\gamma_0=2.4\ eV$ is the nearest neighbor tight binding overlap integral\cite{Reich:2002,Blinowski:1980}). The Hamiltonian, Eq.(\ref{GC2}), corresponds to a two-band model with eigenvalues and eigenvectors given respectively as:
\begin{equation}
E_s=s\sqrt{\Delta^2+(v_0\hbar k)^2}=s E,
\label{GC3}
\end{equation}
\begin{equation}
|{\bf k},s\rangle=\frac{E-s\Delta}{\sqrt{E_0^2+(E-s\Delta)^2}}\left(
\begin{array}{c}
\frac{E_0e^{-i\theta}}{E-s\Delta}\\
 s\\
 \end{array} 
 \right),
\label{GC4}
\end{equation}
where the wave function is
\begin{math}
\psi({\bf r},s)=|{\bf k},s\rangle e^{i{\bf k}\cdot {\bf r}},
\end{math} the two-dimenional wave vector is
 ${\bf k}=k_x{\bf \hat{x}}+k_y{\bf \hat{y}}$ ($\hat{x},\ \hat{y}$ are unit vectors), $e^{i\theta}= (k_x+ik_y)/k$, $E_0=v_0\hbar k$, and $s=\pm$. If $\Delta=0$, one recovers the gapless Dirac-like model suitable for perfect graphene. Using this two-band model and applying linear response theory, the dyadic two dimensional conductivity is found\cite{Falkovsky:2007} to be
 
 \begin{widetext}
\begin{eqnarray}
\label{GC1}
&&\overleftrightarrow{\sigma}(\omega,{\bf q})=\frac{i\hbar e^2}{\pi^2}\times
\sum_{s=\pm}\int d^2k\frac{{\bf v}_{ss}{\bf v}_{ss}\left(f_0[E_s({\bf k}-{\bf q}/2)]-f_0[E_s({\bf k}+{\bf q}/2)]\right)}{[E_s({\bf k}+{\bf q}/2)-E_s({\bf k}-{\bf q}/2)][\hbar\omega-E_s({\bf k}+{\bf q}/2)+E_s({\bf k}-{\bf q}/2)]} +\nonumber \\
&&\frac{2i\hbar^2 e^2\omega}{\pi^2}\int d^2k\frac{{\bf v}_{+,-}{\bf v}_{-,+}\left(f_0[E_+({\bf k}-{\bf q}/2)]-f_0[E_-({\bf k}+{\bf q}/2)]\right)}{[E_-({\bf k}+{\bf q}/2)-E_+({\bf k}-{\bf q}/2)][\hbar^2\omega^2-[E_-({\bf k}+{\bf q}/2)-E_+({\bf k}-{\bf q}/2)]^2]},
\end{eqnarray}
\end{widetext}
\begin{floatequation}
\mbox{\textit{see eq.~\eqref{GC1}}}
\end{floatequation}
where $f_0(E_s)=1/(e^{(E_s-\mu)/k_BT}-1)$ is the equilibrium Fermi distribution function, $k_B$ is the Boltzmann constant and $T$ is the temperature. The velocity matrix elements are ${\bf v}_{ss'}=\langle s,{\bf k}|{\bf\hat{v}}|{\bf k},s'\rangle$, where ${\bf\hat{v}}=\frac{\partial H}{\partial {\bf p}}$. The first term in Eq.(\ref{GC1}) accounts for intraband transitions, while the second one accounts  for interband transitions.  From hence forth, imaginary frequency will be used, which is the relevant quantity for calculating Casimir forces.

\begin{figure}[ht!]
   
    \begin{center}
        \subfigure{
            \label{fig1a}
            \includegraphics[width=0.3\textwidth]{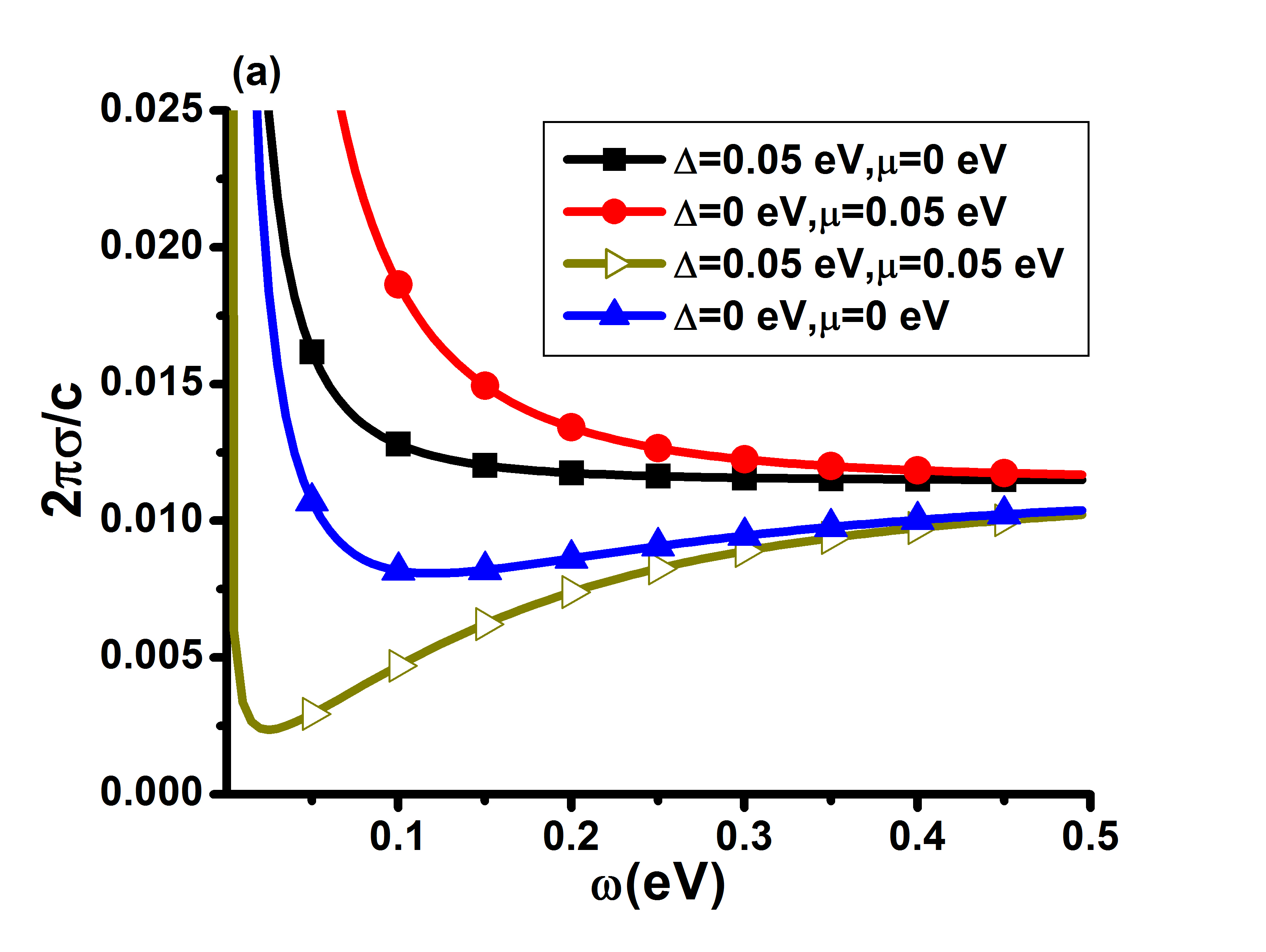}
        }
        
         \subfigure{
            \label{fig1c}
            \includegraphics[width=0.3\textwidth]{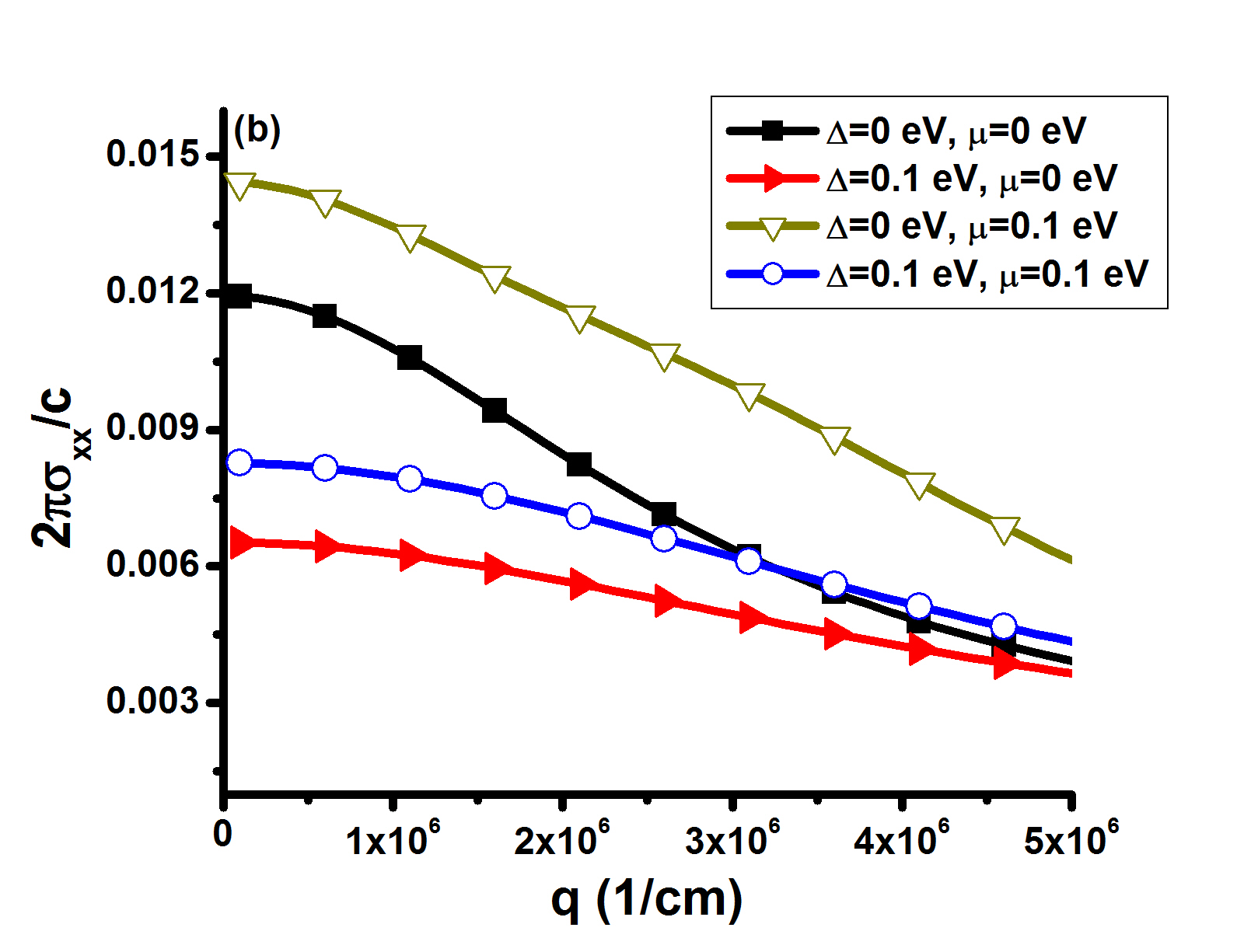}
        }

    \end{center}
    \caption{(a) The isotropic conductivity $\sigma_{0}(i\omega)$ as a function of frequency (in eV) for different values of $\Delta$ and $\mu$ at $T=300$ $K$. (b) $\sigma_{xx}(i\omega)$, as a function of $q$ for finite $\Delta\ \rm{and}\ \mu$ with $\hbar\omega= 0.162\ eV$ corresponding to $T=300\ K$.}
\label{fig1}
\end{figure}

In the limit when $\hbar v_0 q\rightarrow 0$, the optical conductivity of graphene, $\sigma_0(i\omega)$, is isotropic and it is found as:  
\begin{eqnarray}
&&\sigma_{0,intra}(i\omega)=\nonumber \\
&&\frac{e^2\ln(2)}{\hbar^2\pi\omega\beta}+\frac{e^2}{\hbar^2\pi\omega\beta}\ln(\cosh(\Delta\beta)+\cosh(\mu\beta))-\nonumber\\
&&\frac{e^2\Delta^2}{\hbar^2\pi\omega}\int_{\Delta}^\infty\frac{dE}{E^2}\frac{\sinh(E\beta)}{\cosh(\mu\beta)+\cosh(E\beta)},
\label{GC6}
\end{eqnarray}
\begin{eqnarray}
&&\sigma_{0,inter}(i\omega)=\nonumber \\
&&\frac{e^2\omega}{\pi}\int_\Delta^\infty dE\frac{\sinh(E\beta)}{\cosh(\mu\beta)+\cosh(E\beta)}\frac{1}{(\hbar\omega)^2+4E^2}+\nonumber \\
&&\frac{e^2\omega\Delta^2}{\pi}\int_\Delta^\infty\frac{dE}{E^2}\frac{\sinh(E\beta)}{\cosh(\mu\beta)+\cosh(E\beta)}\frac{1}{(\hbar\omega)^2+4E^2},
\label{GC7}
\end{eqnarray}
where $\beta=1/k_BT$.  One notes that when $\mu=0$ and $\Delta=0$, for $k_BT\ll\hbar\omega$, $\sigma_0(i\omega)$ acquires a universal value $\sigma_0(i\omega)\approx \sigma_0\equiv e^2/(4\hbar)$. In Fig(\ref{fig1a}), results are shown for $\sigma_0(i\omega)$ as a function of frequency. The largest differences appear at small $\omega$, where the conductivity is most significantly reduced when $\Delta\neq 0$  ($\mu=0$) and augmented when $\mu\neq 0$ ($\Delta=0$). For larger $\omega$, the conductivity approaches $\sigma_0$.  

If spatial dispersion is taken into account, $\overleftrightarrow{\sigma}$ is not isotropic. For graphene, however, we find that $\sigma_{xx}$ and $\sigma_{yy}$ have very similar magnitude and behavior over large wave vector and frequency regions, therefore only $\sigma_{xx}$ is shown in Fig(\ref{fig1c}). The conductivity, $\sigma_{xx}$, starts changing in a more pronounced way for $q > 10^6\ cm^{-1}$ .
It is noted that $\sigma_{xx}$ is the conductivity in the direction of the electric field parallel to the plane composing graphene sheet while $\sigma_{yy}$ is the direction perpendicular to the electric field.

\section{Graphene Casimir Force and Dispersion\label{DCF}}

The Casimir force per unit area\cite{Abrikosov:1975} between two graphene sheets separated by a distance $d$ is 
\begin{widetext}
\begin{equation}
\label{DCF0}
F=-\frac{ik_BT}{2\pi}\sum_{n=-\infty}^{\infty}
\int_0^\infty h(i|\omega_n|) q dq
\left\{\left[\frac{e^{-2ih(i|\omega_n|)d}}{\rho_{E}(i|\omega_n|)^2}-1\right]^{-1}+
\left[\frac{e^{-2ih(i|\omega_n|)d}}{\rho_{B}(i|\omega_n|)^2}-1\right]^{-1}\right\},
\end{equation}
\end{widetext}
\begin{floatequation}
\mbox{\textit{see eq.~\eqref{DCF0}}}
\end{floatequation}
where $h(i|\omega_n|)=i\sqrt{\epsilon(i|\omega_n|)\mu(i|\omega_n|)(\omega_n/c)^2+q^2}$ with $\omega_n=2\pi nk_BT/\hbar$ and ${\bf q}$ is the two-dimensional wave vector in the $xy$-plane. $\rho_{E,B}$ are the generalized reflection coefficients due to the transverse electric ({\bf E}) and magnetic ({\bf B}) field modes, which are given in terms of the two dimensional conductivity by
\begin{eqnarray}
\rho_E(i\omega)&=&-\frac{2\pi\sigma_{yy}(i\omega,{\bf q})\omega/[h(i\omega)c^2]}{1+2\pi\sigma_{yy}(i\omega,{\bf q})\omega/[h(i\omega)c^2]},\nonumber \\
\rho_B(i\omega)&=&\ \ \ \frac{2\pi\sigma_{xx}(i\omega,{\bf q})h(i\omega)/\omega}{1+2\pi\sigma_{xx}(i\omega,{\bf q})h(i\omega)/\omega}
\label{DCF2}.
\end{eqnarray}

\subsection{{\bf Thermal Limit\label{FSP}}}

The thermal fluctuation forces are first considered, which correspond to the zero Matsubara term ($n=0$) of Eq.(\ref{DCF0}) in the strong dispersion limit, where $\hbar\omega\ll k_BT,\ \hbar v_0 q$.  In most materials purely thermal fluctuation effects are important at distances larger than their characteristic thermal wavelength, $\lambda_T=\hbar c/(k_BT)\sim 7\ \mu m$. Instead, due to the Dirac-like Hamiltonian of graphene and its two dimensional nature, the characteristic thermal wavelength is significantly reduced - $\sim\lambda_T/200$.  As a result, thermal fluctuation forces become important at relatively small scales\cite{GomezSantos:2009}, $d\approx 30\ nm$.  Therefore the quasi-static response is a reasonable approximation for a large distance range. 
 
In the quasi-static response limit of $\hbar  \omega\ll\hbar v_0 q\ll k_BT$, the $\sigma_{xx}$ is 
\begin{equation}
\sigma_{xx}(i\omega,{\bf q})=\frac{\omega q_s}{2\pi q^2}.
\label{FSP1}
\end{equation}
This corresponds to the longitudinal plasmon excitations in the graphene static dielectric function $\epsilon(q)=1+q_s/q$ obtained via the Random Phase Approximation(RPA) method \cite{Ando:2006}. For gapped graphene\cite{Svetovoy:2011},
\begin{eqnarray}
q_s(\Delta)&=&\frac{4\alpha c}{\hbar v_0^2}\left\{2k_BT\ln\left(2\cosh\left(\frac{\Delta}{2k_BT}\right)\right)\right\}
-\nonumber \\
&&\frac{4\alpha c}{\hbar v_0^2}\left\{\Delta\tanh\left(\frac{\Delta}{2k_BT}\right)\right\},
\label{FSP11}
\end{eqnarray}
where $\alpha\approx 1/137$ is the fine structure constant.
Using  Eq.(\ref{DCF0}) and Eq.(\ref{FSP1}) the thermal fluctuation stress between two graphene sheets is obtained to be
\begin{equation}
F(d)=\frac{k_BT}{16\pi d^3}\int_0^\infty\frac{x^2dx}{e^x(x+2dq_s)^2/(2dq_s)^2-1}.
\label{FSP5}
\end{equation}
One finds that as $q_sd\gg1$, the stress approaches the thermal stress between two perfect metals, i.e., $F_T=k_BT\zeta(3)/(8\pi d^3)$.

A similar calculation can be executed for the problem of finding the thermal fluctuation force $f(d)$ between a graphene sheet and a polarizable atom\cite{Caride:2005} with polarizability $\chi(\omega)$.
\begin{widetext}
\begin{equation}
\label{FSP6}
f(d)=-k_BT\sum_{n=-\infty}^{\infty}\chi(i|\omega_n|)\int_0^\infty kdkh^2(i|\omega_n|)\left[2\rho_B(i|\omega_n|)-\left(\frac{\omega_n}{h(i|\omega_n|)c}\right)^2\left[\rho_E(i|\omega_n|)-\rho_B(i|\omega_n|)\right]\right]e^{2idh(i|\omega_n|)}.
\end{equation}
\end{widetext}
\begin{floatequation}
\mbox{\textit{see eq.~\eqref{FSP6}}}
\end{floatequation}
Using Eq.(\ref{FSP1}), the thermal fluctuation force, $n=0$ term, in this case is 
\begin{equation}
f(d)=\frac{k_BT\chi(0)}{8d^4}\int_0^\infty x^3dxe^{-x}\left[\frac{2dq_s}{x+2dq_s}\right],\nonumber
\end{equation}
\begin{eqnarray}
f(d)&=&\frac{k_BT\chi(0)}{8d^4}\left[(2dq_s)^3-(2dq_s)^2+2(2dq_s)\right]-\nonumber \\
&&\frac{k_BT\chi(0)}{8d^4}\left[(2dq_s)^4e^{2dq_s}E_1(2dq_s)\right].
\label{FSP7}
\end{eqnarray}
where $E_{1}(x)$ is the exponential integral. Again as $q_sd\gg 1$, the force between atom and graphene becomes $f_T=3k_BT\chi(0)/(4d^4)$, which is the thermal interaction between an atom and a perfectly conducting plate.

\begin{figure}[ht]
\centering
  \subfigure{
  \label{fig5a}
  \includegraphics[width=0.3\textwidth]{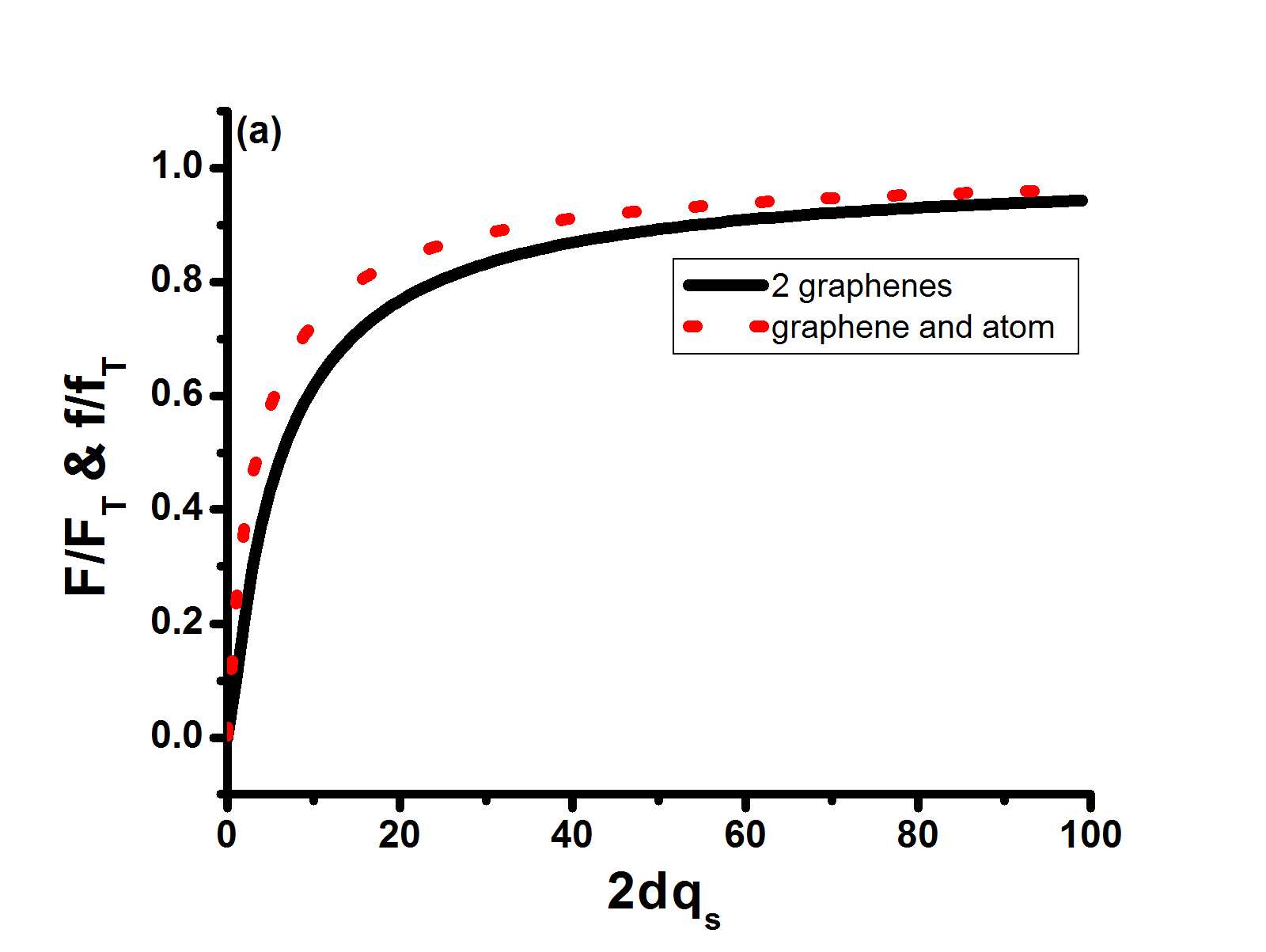}
  }
  
  \subfigure{
  \label{fig5b}
  \includegraphics[width=0.3\textwidth]{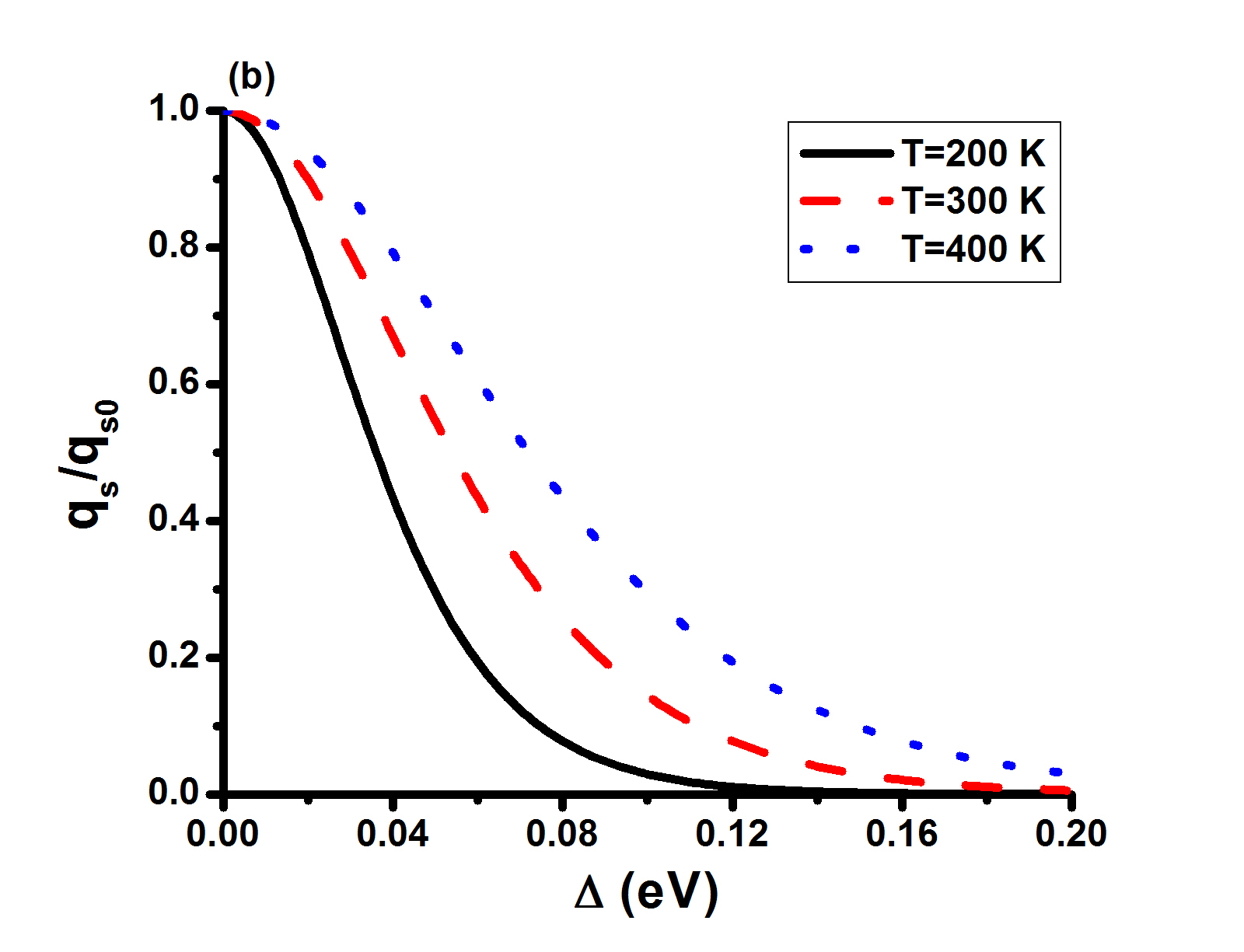}
  }
\caption{(a) Thermal stress between two graphenes normalized to the thermal stress between two metals and thermal force between graphene and atom normalized to the thermal force between metal and an atom. (b) $q_s$ normalized to the value at zero band gap $q_{s0}$ as a function of the energy band gap $\Delta$ for different temperatures, $T=200\ K$, $T=300\ K$, $T=400\ K$.}
\label{fig4}
\end{figure}

The normalized forces,$F/F_T$ and $f/f_T$, as a function of $2dq_s$ are shown in Fig.(\ref{fig4}).  One finds that for large distances spatial dispersion becomes less important, and the force goes to the one expected for ideal metallic surfaces.  For short distances the electron screening shields the force such that it goes to zero at very small distances.  From Fig.(\ref{fig5a}) it is seen that dispersion is significant when $q_sd_\approx 10$ or less.  For the case of gapless graphene, this corresponds to a distance of $d\approx 20\ nm$ at room temperature, in which case higher Matsubara frequencies need to be considered.  Hence, for the case of graphene, dispersion does not play a significant role in the region where the fluctuation force is thermal.
Spatial dispersion becomes important for graphene at larger distances, when $q_s$ decreases.

It has been noted that while for good conductors and dielectrics thermal Van-der-Waal/Casimir forces are not heavily influenced by dispersion effects, the latter may have a strong influence on the force for poor conductors \cite{Pitaevskii:2008,Svetovoy:2008}.  A similar result occurs for graphene, where upon the conduction properties of the material are decreased by introducing a band gap.

By inducing an energy band gap, the effect of spatial dispersion in graphene is increased by effectively decreasing $q_s$, as shown in  Fig.(\ref{fig5b}).
 It is clear that the influence of dispersion increases as $q_s$ gets small with the effect that the thermal fluctuation forces between graphene sheets are substancially decreased.  For example for $\Delta=0.05\ eV$, a typical value for the gap, one obtains a distance for dispersion to have strong effects of $d\approx 40\ nm$, within the range where thermal fluctuation forces are dominant.  Of course, by inducing a sufficiently large band gap, thermal effects cease from becoming dominant.  This effect is analogous to 3-D materials, whereupon the influence of dispersion in the fluctuation forces of a conductor become strong were it to become a poorer conductor.

\subsection{${\bf T=0}$ ${\bf K}$}

When the temperature is zero, the summation in $F$ (Eq.(\ref{DCF0})) is substituted by an integral, $\sum_n=[\hbar/(2\pi k_BT)]\int_0^\infty d\omega$. The $T=0$ $K$ case corresponds to purely quantum mechanical contributions to the Casimir stress. If the graphene conductivity is given by the universal value $\sigma_0=e^2/(4\hbar)$, the graphene/graphene Casimir force per unit area is found as \cite{Dobson:2006,GomezSantos:2009,Drosdoff:2010}
\begin{equation}
F_g=\frac{3e^2}{32\pi d^4}=F_0\kappa(\alpha)
\label{}
\end{equation} 
where $F_0=\pi^2\hbar c/(240 d^{4})$ is the force per unit area between two perfectly conducting metallic sheets and $\kappa(\alpha)=720\alpha/(32\pi^3)$ with $\alpha=1/137$ being the fine structure constant. It is interesting that the force in this limit does not depend on any quantum mechanical characteristics nor the speed of light. Comparing $F_g$ to $F_0$ shows that the distance dependence of the graphene/graphene force is the same as for metallic plates in the retarded regime, however, the magnitude is reduced by an order of the fine structure constant. Incidentally, this differs from the non-retarded ``2D'' metal and doped graphene result--$F\sim d^{-7/2}$\cite{Sernelius:2011,Dobson:2006}

\begin{figure}[ht!]
   
    \begin{center}
        \subfigure{
            \label{fig4a}
            \includegraphics[width=0.3\textwidth]{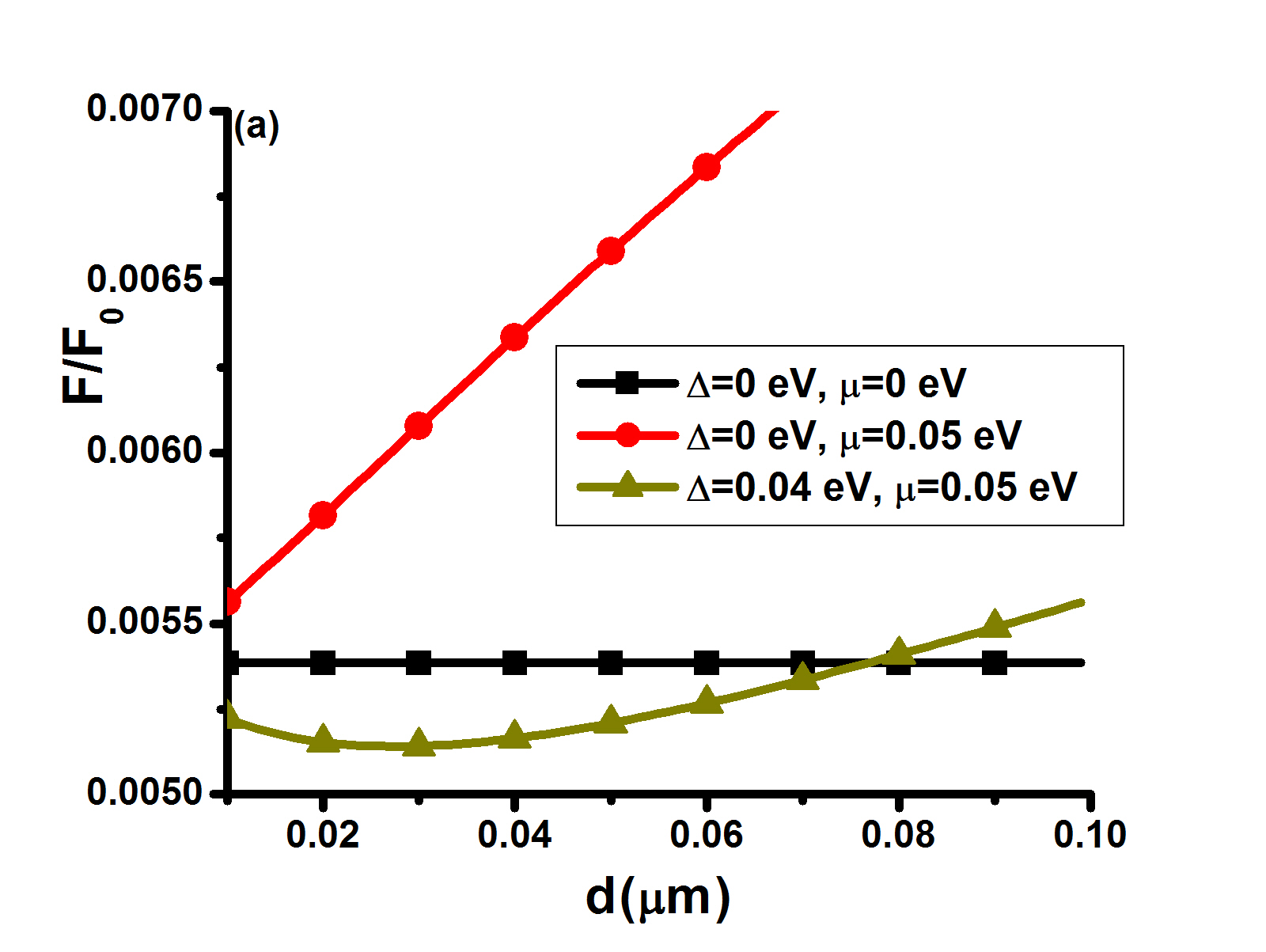}
            }
        \subfigure{
            \label{fig4b}
            \includegraphics[width=0.3\textwidth]{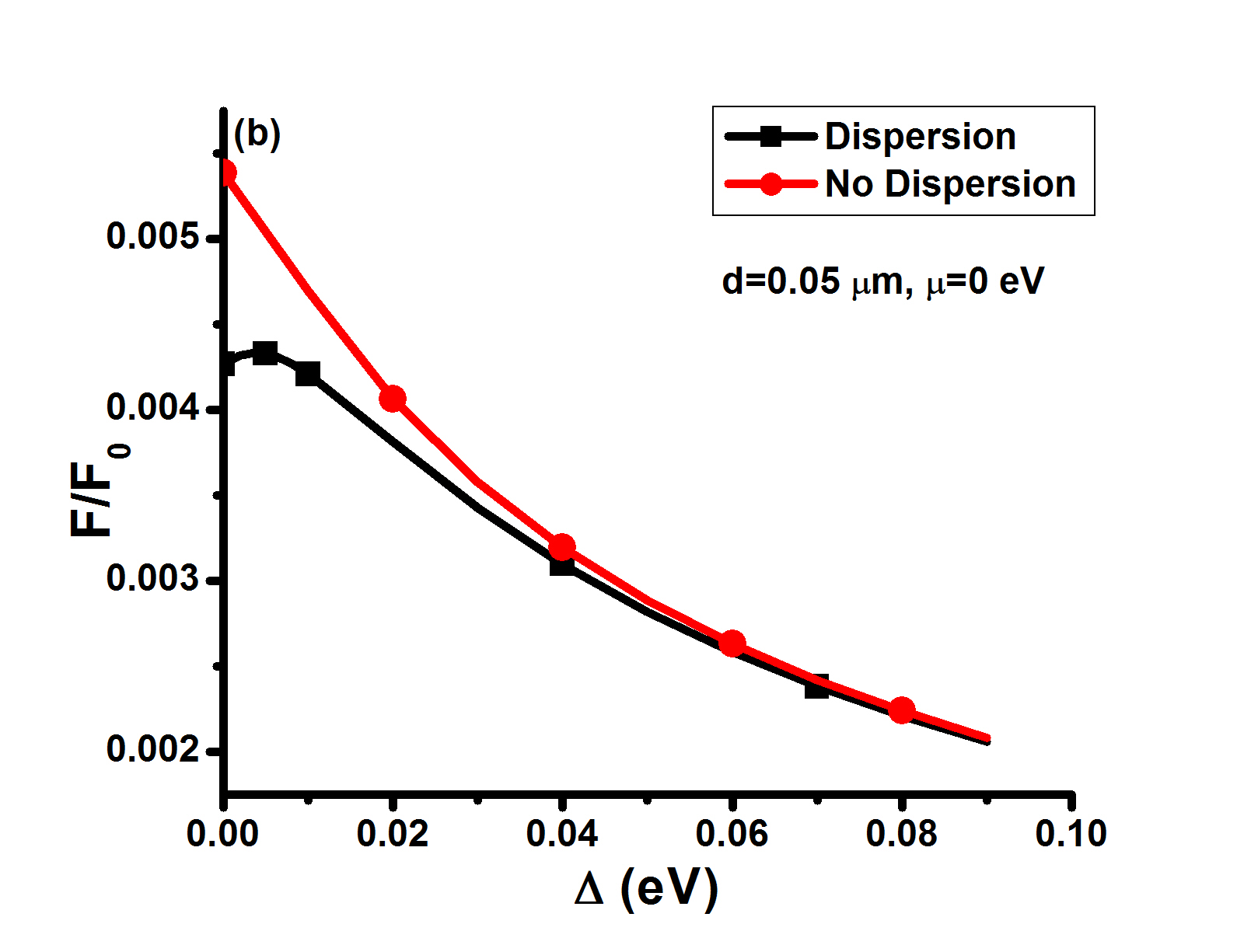}
        }
         \subfigure{
            \label{fig4c}
            \includegraphics[width=0.3\textwidth]{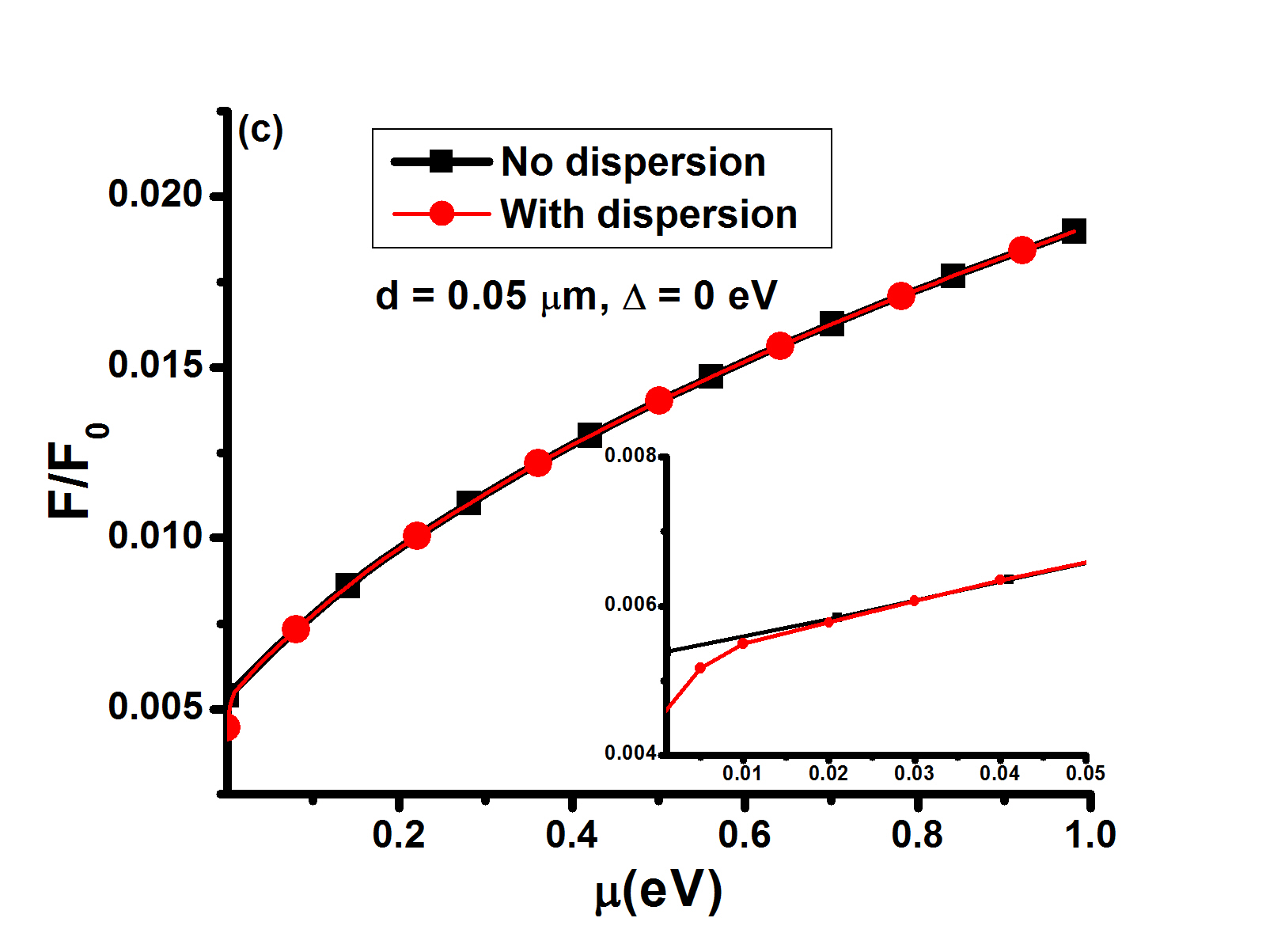}
        }

    \end{center}
    \caption{(a) Casimir stress normalized to $F_0$ at T=0 and calculated using the isotropic conductivity $\sigma_0(i\omega)$ for different $\Delta$ and $\mu$. The flat line indicates the normalized stress obtained via the universal conductivity $\sigma_0$; (b) Casimir stress normalized to $F_0$ at T=0 as a function of $\Delta$ calculated using $\sigma_0(i\omega)$ and $\sigma(i\omega,{\bf q})$ ($\mu=0$); (c) Casimir stress normalized to $F_0$ as a function of $\mu$ calculated using $\sigma_0(i\omega)$ and ${\bf \sigma}(i\omega,{\bf q})$ ($\Delta=0$)  }
\label{fig4d}
\end{figure}

The presence of $\mu$ or $\Delta$ alters the graphene conductivity, which can change the Casimir interaction significantly. Using the isotropic  $\sigma_0(i\omega)$ from Eqs.(\ref{GC6}, \ref{GC7}), it is found that $F/F_0$ increases almost linearly as a function of $d$ when $\mu\neq 0$, while for $\Delta\neq 0$ and $\mu\neq 0$ the normalized Casimir stress has nonlinear dependences - Fig.(\ref{fig4a}). This happens because at shorter distances the effect of the band gap is more pronounced, while at larger distances the finite chemical potential has a stronger effect.  It is found that $F/F_0$ can be diminished by increasing the band gap of the graphene sheets ( Fig.(\ref{fig4b})) or it can be enhanced by increasing the chemical potential ( Fig.(\ref{fig4c})).  These changes can be quite substantial, providing that the system supports large $\mu$ or $\Delta$.

It is interesting to consider to what extent spatial dispersion affects the graphene/graphene Casimir interaction. In the absence of a band gap or chemical potential, dimensional analysis dictates that the force may be written as

\begin{equation}
F=F_0g(\alpha,v_0/c),
\label{DCF2A}
\end{equation}
where the scaling factor $g(\alpha,v_0/c)$ now depends on the fine structure constant, and the ratio of $v_0/c$.  If one uses Eq.(\ref{GC1}), the scaling factor is evaluated as $f(\alpha,v_0/c)\approx 0.0043$.  Note that the distance dependence is still $F\sim d^{-4}$. Comparing the scaling factors $\kappa$ and $g$ shows that the inclusion of spatial dispersion reduces the graphene/graphene Casimir attraction by $\approx 20\ \%$.  If one were to approximate the conductivity of graphene via RPA, then inclusion of dispersion would reduce the graphene/graphene interaction by $\approx 9\ \%$. Figs.(\ref{fig4b},\ref{fig4c}) further show that dispersion effects are rather insignificant as $\mu$ and $\Delta$ increase.

\subsection{${\bf T\neq 0 K}$}

\begin{figure}[ht!]
   
    \begin{center}
        \subfigure{
            \label{fig3b1}
            \includegraphics[width=0.3\textwidth]{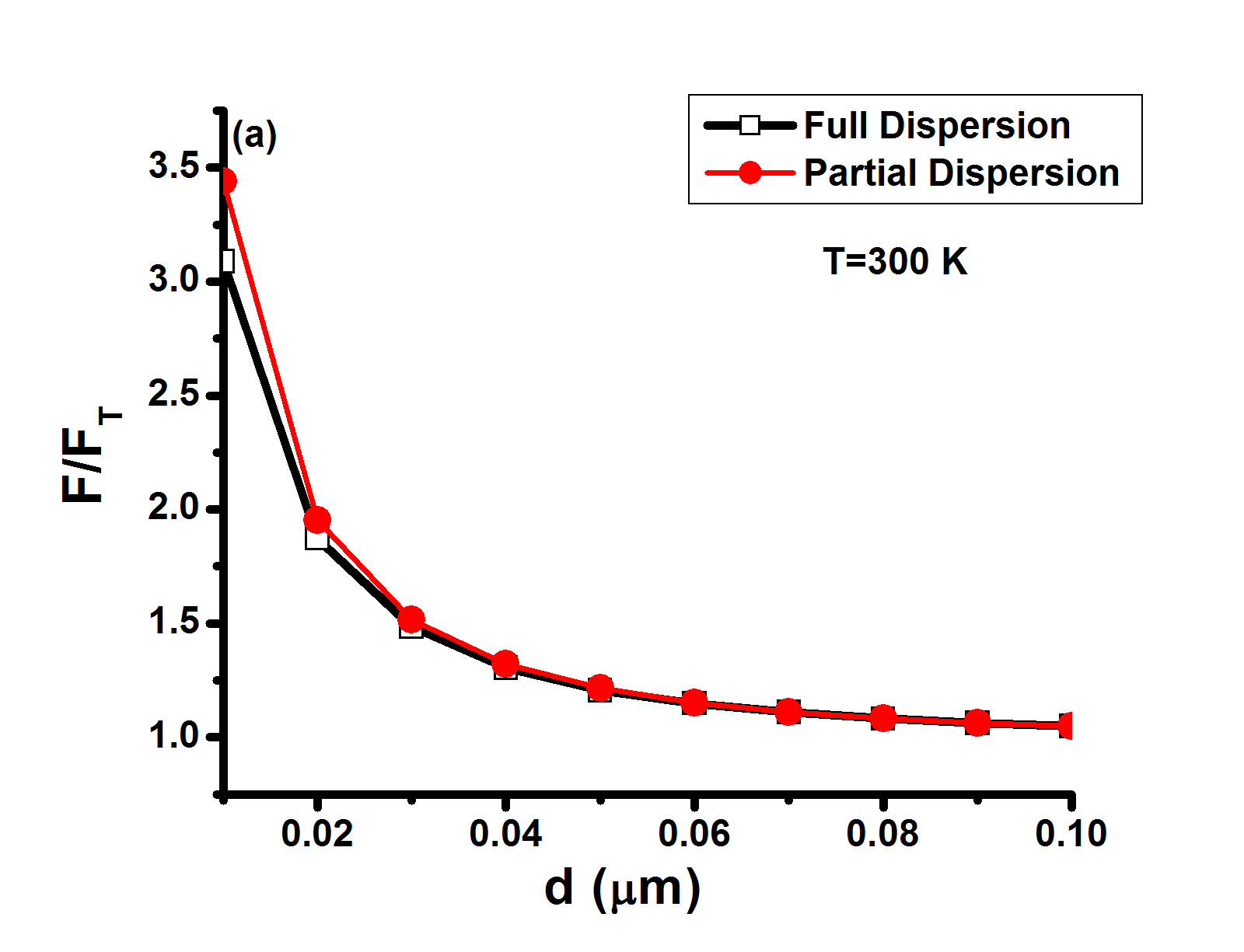}
        } 
        \subfigure{
            \label{fig3b}
            \includegraphics[width=0.3\textwidth]{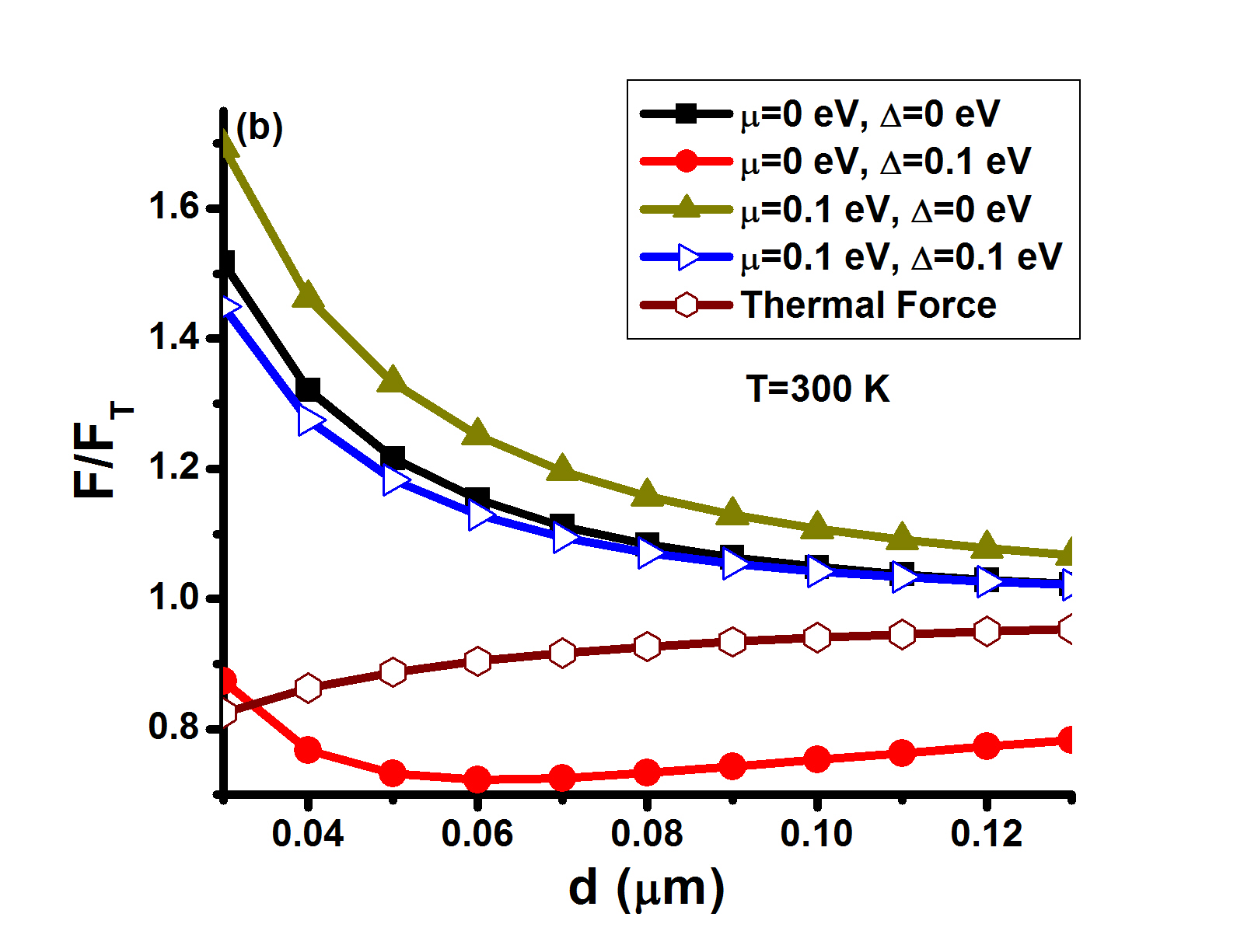}
        }
    \end{center}
    \caption{(a)  Casimir stress normalized to the idealized stress $F_0$ at $T=300\ K$ with and without the inclusion of dispersion for Matsubara terms greater than zero. (b) Casimir stress normalized to the idealized stress $F_0$ at $T=300\ K$.}
\label{fig3}
\end{figure}

Finally, the full finite temperature Casimir interaction is considered. Eq.(\ref{DCF0}) indicates that the largest contribution to the force occurs when $qd\sim 1$.  On the other hand it was shown that for finite frequencies, the conductivity varies sharply as a function of wavenumber when $q_s>10^6\ cm^{-1}$ - Fig(\ref{fig1c}).  Therefore, dispersion effects start becoming important for Matsubara terms in Eq.(\ref{DCF0}) greater than zero at a distance of $d_s\sim 1/q_s=0.01\ \mu m$.  Fig.(\ref{fig3b1}) shows the Casimir interaction force per unit area between two graphene sheets where spatial dispersion is only included in the force for the $n=0$ thermal term and also where spatial dispersion is included for all Matsubara frequency terms.  If the distance is reduced much further the force reduces back to the zero temperature result.

Much more influential parameters to the Casimir interaction are the induced chemical potential or energy band gap.  The effects are displayed in Fig.(\ref{fig3b}), which shows the stress between graphenes for different band gaps, and chemical potentials. In addition, the role of the chemical potential is to increase the magnitude of the force while the energy band gap decreases it (not shown graphically).  The characteristic behavior is similar as for the quantum $T=0$ $K$ limit. 

\section{Conclusion\label{CONC}}

  The Casimir interaction between two graphene sheets has been investigated at room and zero temperatures by using linear response theory with a conductivity response function in order to determine how spatial dispersion affects the interaction. It was found that these effects are not of major importance, either at large or low temperatures.  The Casimir interaction may be well approximated by using a frequency only dependent conductivity. Spatial dispersion corrections start appearing at distances of $d\sim 0.01 \mu m$ at room temperature.  A finite chemical potential can enhance the Casimir interaction substantially and further inhibit spatial dispersion. If a band gap is induced at finite temperature, however, graphene exhibits similar characetristics to a poor metal and dispersion effects start becoming more important, that is the thermal fluctuation forces are reduced by the presence of electronic charge screening. 
  
In the present work, dispersion has been investigated using the 2-band model. Yet
 at shorter distances, in the order of $10\ {\rm nm}$ or less, higher band transitions become important for the Van-der Waals/Casimir interaction.  The inclusion of such effects without spatial dispersion have already been investigated in a previous work\cite{Drosdoff:2011}.   
   
\section{Acknowledgements}

We acknowledge financial support from the Department of Energy under contract DE-FG02-06ER46297.  I. V. B. acknowledges support from NSF-HRD-0833184 and ARO-W911NF-11-1-0189.


\begin{thebibliography}{99}

\bibitem{Casimir:1948}
\Name{Casimir H.~B.~G.}
\REVIEW{Proc. K. Ned. Akad. Wet.}{51}{1948}{793}.
\bibitem{London:1930}
\Name{F.~London}
\REVIEW{Z. Phys.}{63}{1930}{245}.
\bibitem{Abrikosov:1975}
\Name{Abrikosov A.~A., Gorkov L.~P., \and Dzyaloshinski I.~E.}
\Book{ Methods of Quantum Field Theory in Statistical Physics}
\Publ{ Dover Publications, Inc., New York}
\Year{1975}.
\bibitem{Rodriguez:2011}
\Name{Rodriguez A.~W., Capasso Federico and Johnson Steven G.}
\REVIEW{nature photonics}{5}{2011}{211}.
\bibitem{Novoselov:2004}
\Name{Novoselov K.~S., Geim A.~K.~, Morozov S.~V.~, Jiang D., Zhang Y., Dubonos S.~V., Grigorieva I.~V., and Firsov A.~A.}
\REVIEW{Science}{360}{2004}{666}. 
\bibitem{Novoselov:2005}
\Name{Novoselov K.~S., Jiang D., Schedin F., Booth T.~J., Khotkevich V.~V., Morozov  S.~V., and Geim A.~K.}
\REVIEW{Proc. Nat. Acd. Sci.}{102}{2005}{10451}.
\bibitem{Lin:2010}
\Name{Lin Y.~M., Dimitrakopoulos C., Jenkins K.~A., Farmer D.~B., and Chiu H.~Y.}
\REVIEW{Science}{327}{2010}{662}.
\bibitem{Stoller:2008}
\Name{Stoller M.~D., Park S., Zhu Y., An J., and Ruoff R.~S.}
\REVIEW{Nano Lett.}{8}{2008}{3498}.
\bibitem{Dobson:2011}
\Name{Dobson J. F.}
\REVIEW{Surf. Sc.}{605}{2011}{1621}.
\bibitem{Drosdoff:2011}
\Name{Drosdoff D. and Woods L. M.}
\REVIEW{Phys. Rev. A}{84}{2011}{062501}.
\bibitem{Sarabadani:2011}
\Name{Sarabadani J.~l , Naji A., Asgari R., and Podgornik R.}
\REVIEW{Phys. Rev. B}{84}{2011}{155407}.
\bibitem{Svetovoy:2011}
\Name{Svetovoy V., Moktdadir Z., Elwenspoek M. and Mizuta H.}
\REVIEW{EPL}{96}{2011}{14006}.
\bibitem{Fialkovsky:2011}
\Name{Fialkovsky I.~V., Marachevsky V.~N., and Vassilevich D.~V.}
\REVIEW{Phys. Rev B}{84}{2011}{035446}.
\bibitem{Sernelius:2011}
\Name{Sernelius B.~E.}
\REVIEW{Europhys. Lett.}{95}{2011}{57003}.
\bibitem{Drosdoff:2010}
\Name{Drosdoff D. and Woods L.~M.}
\REVIEW{Phys. Rev. B}{82}{2010}{155459}.
\bibitem{GomezSantos:2009}
\Name{G\'omez-Santos G.}
\REVIEW{Phys. Rev. B}{80}{2009}{245424}.
\bibitem{Dobson:2006}
\Name{Dobson J.~F., White A., and Rubio A.}
\REVIEW{Phys. Rev. Lett}{96}{2006}{073201}.
\bibitem{Sernelius:2006}
\Name{Sernelius B.~E.}
\REVIEW{J. Phys. A:Math. Gen.}{39}{2006}{6741}.
\bibitem{Esquivel:2006}
\Name{Esquivel-Sirvent R., Villarreal C., Moch\'an W.~L., Contreras-Reyes A.~M., and Svetovoy V.~B.}
\REVIEW{J. Phys. A: Math. Gen.}{39}{2006}{6323}.
\bibitem{Brenner:2012}
\Name{Brenner K., Yang Y., and Murali R.}
\REVIEW{Carbon}{50}{2012}{637}.
\bibitem{Lin:2012}
\Name{Lin Z., Song M., Ding D., Liu Y., Liu M., and Wong C.}
\REVIEW{Phys. Chem. Chem. Phys.}{14}{2012}{3381}. 
\bibitem{Zhou:2007}
\Name{Zhou S.~Y., Gweon G.~-H., Fedorov, A.~V., First P,~N.~, De Heer W.~A., Lee D.~-H., Guinea F., Castro Neto A.~H., and Lanzara A.}
\REVIEW{Nature Materials}{6}{2007}{770}.
\bibitem{Yavari:2010}
\Name{Yavari F., Kritzinger C., Gaire C., Song L., Gullapalli H., Borca-Tasciuc T., Ajayan P.~M., Koratkar N.}
\REVIEW{Small}{6}{2010}{2535}.
\bibitem{Haberer:2010}
\Name{Haberer D., Vyalikh D.~V., Taioli S., Dora B., Farjam M., Fink J., Marchenko D., Pichler T., Ziegler K., Simonucci S., Dresselhaus M.~S., Knupfer, B\"{u}chner, and Gr\"{u}neis}
\REVIEW{Nano Lett.}{10}{2010}{3366}.
\bibitem{Falkovsky:2007}
\Name{Falkovsky L. A., \and Varlamov A. A.}
\REVIEW{Eur. Phys. J. B}{56}{2007}{281}.
\bibitem{Kubo:1957}
\Name{Kubo R.}
\REVIEW{J. of the Phys. Soc. of Jap.}{12}{1957}{570}.
\bibitem{Reich:2002}
\Name{Reich S., Maultzsch J., \and Thomsen C.}
\REVIEW{Phys. Rev. B}{66}{2002}{0035412}.
\bibitem{Blinowski:1980}
\Name{Blinowski J., Nguyen Hy Hau, Rigaux C., Vieren J.~P., Le Toullec R., Furdin G., H\'erold A., and Melin J.}
\REVIEW{J.~Physique}{41}{1980}{47}.
\bibitem{Ando:2006}
\Name{Ando T.}
\REVIEW{J. of the Phys. Soc. of Jap.}{75}{2006}{074716}.
\bibitem{Caride:2005}
\Name{Caride A.~O., Klimchitskaya G.~L., Mostepanenko V.~M., and Zanette S.~I.} \REVIEW{Phys. Rev. A}{71}{2005}{042901}.
\bibitem{Pitaevskii:2008}
\Name{Pitaevskii L.~P.}
\REVIEW{Phys. Rev. Lett.}{101}{2008}{163202}.
\bibitem{Svetovoy:2008}
\Name{Svetovoy V.~B.}
\REVIEW{Phys. Rev. Lett.}{101}{2008}{163603}.


\end{thebibliography}
\end{document}